\documentclass[twocolumn,showpacs,amsfonts,aps,prc,floatfix,%
superscriptaddress]{revtex4} 

\usepackage{amsmath}
\usepackage{bm}
\usepackage{graphicx}

\newcommand{\be}[1]{\begin{equation}\label{#1}}
\newcommand{\ee}{\end{equation}}

\newcommand{\eq}{{\,=\,}}

\usepackage{color}

\begin{document}


\title{Eccentricity fluctuation effects on elliptic flow
in relativistic heavy ion collisions}
\date{\today}

\author{Tetsufumi Hirano}
\email{hirano@phys.s.u-tokyo.ac.jp}
\affiliation{Department of Physics, The University of Tokyo,
Tokyo 113-0033, Japan}

\author{Yasushi Nara}
\email{nara@aiu.ac.jp}
\affiliation{Akita International University, Yuwa, Akita-city, 010-1292, Japan}

\begin{abstract}
We study effects of eccentricity fluctuations
on the elliptic flow coefficient $v_2$ at mid-rapidity
in both Au+Au and Cu+Cu collisions at $\sqrt{s_{NN}}=200$\,GeV
by using a hybrid model that combines ideal hydrodynamics for
space-time evolution of the quark gluon plasma phase and a
hadronic transport model for the hadronic matter.
For initial conditions in hydrodynamic simulations,
both the Glauber model and the color glass condensate
model are employed to demonstrate the effect of
initial eccentricity fluctuations originating from the nucleon position inside
a colliding nucleus.
The effect of eccentricity fluctuations is modest in semicentral
Au+Au collisions, but significantly enhances $v_2$ in Cu+Cu collisions.
\end{abstract}

\maketitle

\section{Introduction}
\label{sec1}
\vspace*{-2mm}

One of the major discoveries at 
Relativistic Heavy Ion Collider (RHIC)
in Brookhaven National Laboratory (BNL)
is that azimuthally anisotropic flow (the so-called
elliptic flow) \cite{Ollitrault}
is found to be as large as an ideal hydrodynamic prediction
for the first time
in relativistic heavy ion collisions \cite{STARv2,PHENIXv2,reviews}.
Whether local thermalization,
which is demanded in application of hydrodynamics,
is reached is not known a priori
in relativistic heavy ion collisions.
Therefore the discovery indicates
that the heavy ion reactions at relativistic energies
provide a unique opportunity
to investigate the high temperature
QCD matter \textit{in equilibrium}.
The discovery also indicates a new possibility to obtain the transport
properties of the QCD matter under extreme conditions.
Effects of viscosity have already been taken into account
in several hydrodynamic simulations 
\cite{Song,Chaudhuri,Baier,Dusling,Denicol}
and also specifically in distortion of distribution functions
\cite{Teaney,Monnai}
to calculate elliptic flow coefficients.

Systematic studies showed, however, that the reasonable agreement
between results from
ideal hydrodynamics and elliptic flow data
has been achieved only by a particular combination of dynamical modeling,
namely
initial conditions from the Glauber model,
the perfect fluid quark gluon plasma (QGP) core, and 
the dissipative hadronic corona \cite{HG05}.
For example, when one takes
 initial conditions
from the color glass condensate (CGC) model
instead of the conventional Glauber model ones,
the result overshoots
the elliptic flow data  \cite{HHKLN}
due to initial eccentricity larger than that from the
Glauber model \cite{Kuhlman:2006qp}.
Even within the Glauber model initializations,
the agreement between hydrodynamic results and
the data, in particular centrality dependence of the elliptic flow,
is not perfect, which would be due to an absence of
initial fluctuation effects \cite{MS03,
Zhu:2005qa,Bhalerao:2006tp,phobosfluc,Andrade:2006yh,Drescher:2006ca,Broniowski:2007ft,Broniowski:2007nz,Voloshin:2007pc}.
Therefore, further investigation is indispensable toward better understanding
of the elliptic flow data and, in turn, understanding of
transport properties of the QGP.

In this paper, we follow up with the previous study \cite{HHKLN,HHKLN2}
based on an ideal hydrodynamic description of the QGP fluid and
a microscopic description of the hadronic gas
by taking into account the initial eccentricity fluctuation
effects.
It was found that large initial eccentricity
in some particular modeling of the CGC picture
is attributed to improper treatment
of nuclear edge regions to define the saturation scale \cite{Drescher:2006ca}.
Improved treatment of the edge
leads to slight reduction of eccentricity \cite{Drescher:2006ca}.
So it would be interesting to see whether the improved CGC model 
ends up with reproduction
of elliptic flow data as well as how initial
fluctuation of eccentricity
affects elliptic flow coefficients 
in both the Glauber model and the CGC model.

This paper is organized as follows.
In Sec.~\ref{sec2},
we first overview the dynamical modeling
of relativistic heavy ion collisions
based on a hybrid (hydro+cascade) approach.
We discuss how to implement the eccentricity
fluctuation in the hydrodynamic
initial conditions.
For the initial transverse profiles,
we employ the Glauber model and the CGC
model and compare the eccentricities with each
other.
In Sec.~\ref{sec3},
we investigate the effect of
eccentricity fluctuation on the elliptic flow
coefficients
by using the hybrid dynamical model.
Section \ref{sec4} is devoted to conclusion.

\section{The model}
\label{sec2}
\subsection{Dynamical modeling of heavy ion collisions}

Our study is mainly based on a hybrid model which combines an ideal fluid 
dynamical description of the QGP stage with a realistic kinetic 
simulation of the hadronic stage 
\cite{HHKLN,HHKLN2,BassDumitru,TLS01,NonakaBass,Petersen:2008dd,Werner}.
Relativistic hydrodynamics 
is the most relevant framework to understand the bulk and transport 
properties of the QGP since it directly connects the collective flow 
developed during the QGP stage with its equation of state (EOS).
It is based on the key assumption of local thermalization. Since this
assumption breaks down during both the very anisotropic initial matter 
formation stage and the dilute late hadronic rescattering stage, the
hydrodynamic framework can be applied at best 
only during the intermediate period.
To describe the breakdown of the hydrodynamic
description
during the late hadronic stage due to expansion and dilution of the 
matter, one may have two options: One can either impose a sudden transition
from thermalized matter to non-interacting
and free-streaming hadrons through
the Cooper-Frye prescription \cite{CF} at 
a decoupling temperature $T_{\mathrm{dec}}$,
or make a transition from a macroscopic 
hydrodynamic description to a microscopic kinetic description at 
a switching temperature $T_{\mathrm{sw}}$.
We here use the first approach
to fix the initial parameters by comparison with the multiplicity data.
After all initial parameters are fixed, we carry out
hybrid simulations to investigate the effect of initial
eccentricity fluctuations as well as 
hadronic dissipation. 

For the space-time evolution of the perfect QGP fluid we solve numerically
the equations of motion of ideal hydrodynamics for a given initial state
in three spatial dimensions and in time
\cite{Hiranov2eta}:
\begin{eqnarray}
\label{eq:hydro}
\partial_\mu T^{\mu \nu} & = & 0, \\
\label{eq:tmunu}
T^{\mu \nu} &  =  &(e+p) u^\mu u^\nu -p g^{\mu \nu}.
\end{eqnarray}
Here $e$, $p$, and $u^\mu$ are energy density, pressure, and four-velocity
of the fluid, respectively. 
We neglect the net baryon density
due to its smallness at collider energies~\cite{BRAHMS:stopping}.
We have solved Eq.~(\ref{eq:hydro}) in $(\tau,\,x,\,y,\,\eta_s)$
coordinates \cite{Hiranov2eta} where $\bm{x}_\perp = (x,y)$
is a transverse coordinate,
$\tau\eq\sqrt{t^2{-}z^2}$ and 
$\eta_s\eq\frac{1}{2}\ln[(t{+}z)/(t{-}z)]$ are longitudinal proper time 
and space-time rapidity, respectively, adequate for the description of 
collisions at ultra-relativistic energies.
As will be discussed later in the next subsection,
we calculate the initial condition only at midrapidity
and neglect possible fluctuation and correlation effects
in the forward/backward rapidity regions in this work.
So we assume longitudinal boost invariance up to the longitudinal boundary of
the mesh at $\eta_s = 5$
in three dimensional grids
and calculate the observables only at midrapidity. 
The solution for
the positive rapidity region is properly reflected
to the negative rapidity region.
We have checked that there is neither rapidity dependence
nor numerical artifacts from the finite volume in the longitudinal direction
on the observables at midrapidity.

For the high temperature QGP phase ($T>T_c=170$ MeV) we use the
EOS of massless parton gas 
($u$, $d$, $s$ quarks and gluons) with a bag pressure $B$:
\begin{eqnarray}
  p = \frac{1}{3}(e-4B)
\end{eqnarray}
The bag constant is tuned to $B^{\frac{1}{4}} = 247.19$\,MeV to ensure
a first order phase transition to a hadron resonance gas at critical
temperature $T_c = 170$\,MeV. The hadron resonance gas model at
$T<T_c$ includes all hadrons up to the mass of the $\Delta(1232)$
resonance.
Our hadron 
resonance gas EOS implements chemical freeze-out at 
$T_\mathrm{chem}{\eq}T_c = 170$ MeV \cite{HT02}.
This is achieved by introducing appropriate 
temperature-dependent chemical potentials $\mu_i(T)$ for each hadronic 
species $i$ \cite{HT02,Bebie,spherio,Teaney:2002aj,Kolb:2002ve,Huovinen:2007xh}.
In our hybrid model simulations we switch from ideal hydrodynamics to 
a hadronic cascade model at the switching temperature 
$T_{\mathrm{sw}} = 169$\,MeV. The subsequent hadronic rescattering 
cascade is modeled by JAM \cite{jam}, initialized with hadrons 
distributed according to the hydrodynamic model output and calculated
with the Cooper-Frye formula \cite{CF} along the $T_{\mathrm{sw}}\eq169$\,MeV 
hypersurface rejecting inward-going particles.

\subsection{Initial conditions with eccentricity fluctuation effects}

The space-time evolution of thermodynamic variables
is described in the hydrodynamic framework.
The concept of ensemble average
in a sense of statistical mechanics is implicitly
there to interpret the thermodynamic variables.
So we identify the ensemble with
a large number of collision events 
and suppose the hydrodynamic solution
represents an average behavior
of the space-time evolution of the matter
for a given centrality cut.
This is contrary to an approach in 
Refs.~\cite{Andrade:2006yh,Andrade:2008xh} 
in which hydrodynamic equations 
with lumpy initial conditions are solved
in an event-by-event basis.
The smooth initial condition 
is, however, required in some practical reasons
in our approach.
Such a lumpy initial condition could 
generate sizable numerical viscosity
in hydrodynamic simulations if the mesh
size is not sufficiently small.
In the present study, the mesh size in the transverse
direction is $\Delta x = 0.3$ (0.2) fm in Au+Au (Cu+Cu) collisions.
The mesh size is of the same order of the transverse
size of nucleons, so it would be hard to capture possible lumpy structures
in the initial conditions.
Moreover, one needs to perform a large number of
simulations to gain sufficient statistics
in final observables.
Therefore
we pursue a more conservative approach by initializing
the distribution of thermodynamic variables
with smooth functions averaged over many events.

For initial conditions in this study, we employ 
the Monte-Carlo version of both the Glauber model and
the factorized Kharzeev-Levin-Nardi (fKLN) model \cite{Drescher:2006ca}
to generate the initial distribution of
entropy density
in an event-by-event basis.
One has extensively used the Monte Carlo Glauber model (MC-Glauber)
to determine
the centrality cut, the average numbers
of participants and binary collisions for a given centrality,
and so on \cite{Miller:2007ri}. 
On the other hand, the fKLN model enables us 
to improve the treatment of entropy production processes.
This model also gives a natural
description near the edge regions compared to
the ordinary KLN approach 
\cite{KLN01}
employed by us previously \cite{Hirano:2004rs}.
The Monte-Carlo version of fKLN,
which we call MC-KLN, implements
the fluctuations of gluon distribution
due to the position of hard sources 
(nucleons) in the transverse plane \cite{Drescher:2006ca}.

We first calculate a transverse entropy density profile
\begin{equation}
s_0(\bm{x}_{\perp}) = s(\tau = \tau_0, x, y, \eta_s = 0)
\end{equation}
in each sample
at an impact parameter $b$ for a given centrality,
where $\tau_0=0.6$ fm/$c$ is the initial time for the hydrodynamical
simulations.
Then  the variances of the profile are obtained from
\begin{eqnarray}
\sigma_x^2 & = & \langle x^2 \rangle - \langle x \rangle^2,\\
\sigma_y^2 & = & \langle y^2 \rangle - \langle y \rangle^2,\\
\sigma_{xy} & = & \langle xy \rangle - \langle x \rangle\langle y \rangle.
\end{eqnarray}
Here $\langle \cdots \rangle$ describes the average
over transverse plane by weighting the entropy density in a single sample:
\begin{equation}
\langle \cdots \rangle = \frac{\int d^2 x_\perp \cdots s_0(\bm{x}_\perp)}{\int d^2 x_\perp s_0(\bm{x}_\perp)}.
\end{equation}
The eccentricity with respect to reaction plane,
the participant eccentricity,
and the corresponding transverse areas
can be defined \cite{phobosfluc}, respectively, as
\begin{eqnarray}
\varepsilon_{\mathrm{RP}} & = & 
  \frac{\sigma_y^2- \sigma_x^2}{\sigma_y^2+ \sigma_x^2},\\
\varepsilon_{\mathrm{part}} & = & \frac{\sqrt{(\sigma_y^2- \sigma_x^2)^2+4\sigma_{xy}^2}}{\sigma_y^2+ \sigma_x^2},\\
S_{\mathrm{RP}} & = & \pi \sqrt{\sigma_x^2 \sigma_y^2}, \\
S_{\mathrm{part}} & = & \pi \sqrt{\sigma_x^2 \sigma_y^2-\sigma_{xy}^2}. 
\label{eq:area}
\end{eqnarray}
It should be noted here that,
in calculating eccentricity and transverse area,
we use the entropy density profile instead of
the distribution of participants.
Thus the eccentricity depends on
modeling of initialization, \textit{i.e.}, transverse
profiles in hydrodynamic simulations.

The impact parameter vector and the true reaction plane
are not known experimentally.
So one can set an apparent frame of created matter shifted 
by $(x, y) = (\langle x \rangle, \langle y \rangle)$
and tilted by $\Psi$ from
the true frame in the transverse plane \cite{phobosfluc}:
\begin{eqnarray}
\tan 2 \Psi = \frac{\sigma_y^2-\sigma_x^2}{2\sigma_{xy}}.
\end{eqnarray}
The anisotropy of particle distribution
could be correlated with the misidentified frame.
To take account of this,
we first shift the center-of-mass of the system
to the origin in the calculation frame and then
rotate the profile in the azimuthal direction by $\Psi$
to match the apparent reaction plane to
the true reaction plane.
We generate the next sample of an entropy density profile again as above
and \textit{average the profiles over many samples}.
We repeat the above procedure for many samples
until the initial distribution is smooth enough.
The initial conditions obtained in this way 
contain the effects of eccentricity fluctuation
even though the profile is smooth.
In particular, even in case of vanishing impact parameter,
eccentricity is finite due to its event-by-event fluctuation.
It is the particle distribution calculated from the initial conditions
mentioned above that can be directly compared with the experimental data. 
Note that the procedure
averaging over many samples without shift or rotation
corresponds to conventional initialization
without the effect of
eccentricity fluctuation.

\subsection{Monte Carlo Glauber model}

In the Monte Carlo version of the Glauber model,
we first sample the positions of nucleons
according to a nuclear density distribution
for two colliding nuclei.
A nucleon-nucleon collision takes place if their distance $d$ in the
transverse plane (orthogonal to the beam axis) satisfies the condition
\begin{equation}
 d \leq \sqrt{\sigma_{NN}^{\text{in}}/\pi},
\end{equation}
where $\sigma_{NN}^{\text{in}}=41.94$ mb is the inelastic nucleon-nucleon
cross section at $\sqrt{s_{NN}}=200$ GeV.
The number of binary collisions $N_{\mathrm{coll}}$
is obtained by counting total number of nucleon-nucleon collisions,
and the number of participants $N_{\mathrm{part}}$
is the total number of nucleons which collide at least once.

\begin{figure}[htb]
\includegraphics[width=3.4in]{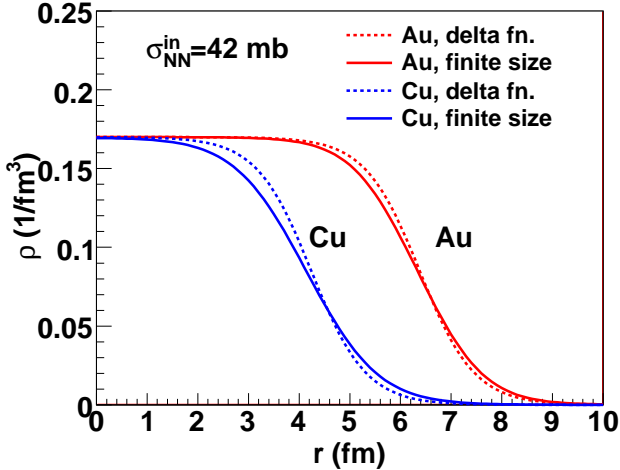}
\caption{(Color online)
Nuclear density as a function of nuclear radius.
Solid lines show nuclear density distribution for gold and copper
nuclei in which a finite nucleon profile 
is implemented and positions of nucleons 
are sampled according to the Woods-Saxon distribution
with default parameter sets.
Dashed lines show the Woods-Saxon distribution with default parameter sets. 
}
\label{fig:WS}
\end{figure}

In the ordinary MC-Glauber model, the following Woods-Saxon distribution
is employed as a nuclear density profile 
\begin{eqnarray}
\label{eq:WS}
\rho_{\mathrm{WS}} & = & \frac{\rho_0}{\exp\left(\frac{r-r_0}{d} \right)+1},
\end{eqnarray}
where $\rho_0 =$ 0.1695 (0.1686) fm$^{-3}$, $r_0  = $ 6.38 (4.20641) fm, 
$d  = $ 0.535 (0.5977) fm for a gold (copper) nucleus 
\cite{Miller:2007ri,De Jager:1987qc}.
If one assumes a profile of nucleon as the delta function,
the resulting nuclear density is nothing but
the Woods-Saxon distribution
\begin{eqnarray}
\rho(\vec{x}) = \int \delta^{(3)}(\vec{x}-\vec{x}_0)\rho_{\mathrm{WS}}(\vec{x}_0)d^3\vec{x}_0 .
\end{eqnarray}
However, in the case of finite size profile for nucleons
as assumed in the MC-Glauber model,
the nuclear density is no longer the same as the Woods-Saxon distribution.
\begin{eqnarray}
\rho(\vec{x}) & = &\int \Delta(\vec{x}-\vec{x}_0)\rho_{\mathrm{WS}}(\vec{x}_0)d^{3}\vec{x}_0, \\
\label{eq:finiteprofile}
\Delta(\vec{x}-\vec{x}_0) & = & \frac{\theta(r_{N} -\mid \vec{x}-\vec{x}_0 \mid)}{V_{N}},\\
V_{N} & = & \frac{4 \pi r_{N}^3}{3}, \quad 
r_{N}  =  \sqrt{\frac{\sigma_{NN}^{\mathrm{in}}}{\pi}}.
\end{eqnarray}
This is illustrated in Fig.~\ref{fig:WS}.
%
Obviously, nuclear surface is more diffused
in both gold and copper nuclei
due to the finite nucleon profile in Eq.~(\ref{eq:finiteprofile}).
As a matter of fact,
eccentricity becomes smaller by $\sim 10\%$ \cite{Drescher:2006ca}
unless one adjusts the Woods-Saxon parameters according to the finite nucleon
profile.
So we parameterize the distribution of nucleon positions
to reproduce the Woods-Saxon distribution 
with default parameters in Eq.~(\ref{eq:WS}) \cite{Broniowski:2007nz}.
In our MC-Glauber model,
we find the default Woods-Saxon distribution
is reproduced by 
a larger radius parameter $r_0  = $ 6.42 (4.28) fm and 
a smaller diffuseness parameter 
$d  = $ 0.44 (0.50) fm
(\textit{i.e.}, sharper
boundary of a nucleus)
for a gold (copper) nucleus
than the default parameters.
This kind of effect exists almost all Monte Carlo approaches
of the collisions
including event generators.
If one wants to discuss eccentricity and elliptic flow coefficient $v_2$
within $\sim 10\%$ accuracy, this effect should be taken into account.

We assume that the initial entropy profile
in the transverse plane at midrapidity is proportional
to a linear combination of the number density
of participants and that of binary collisions in the Glauber model
\begin{eqnarray}
s_0(\bm{x}_\perp) 
& = & \left. \frac{dS}{\tau_0 dx dy d\eta_s} \right|_{\eta_s = 0}  \nonumber\\
& = & \frac{C}{\tau_0}\left(\frac{1-\delta}{2} \frac{dN_{\mathrm{part}}}{d^2x_{\perp}} + \delta \frac{dN_{\mathrm{coll}}}{d^2x_{\perp}}\right).
\end{eqnarray}
Parameters $C=19.8$ and $\delta=0.14$ have been fixed through
comparison with the centrality dependence of multiplicity data
in Au+Au collisions at RHIC \cite{PHOBOS_Nch} by pure hydrodynamic calculations
with $T_{\mathrm{dec}}=100$\,MeV.
Notice that the total multiplicity of the produced
particle does not change during the late hadronic stage
due to the chemical freezeout in the EOS.

\subsection{Monte Carlo KLN model}

In the MC-KLN model,
the saturation scale
for a nucleus $A$
at a transverse coordinate $\bm{x}_{\perp}$ in each sample is given by
\begin{eqnarray}
Q_{s,A}^2 (x; \bm{x}_\perp) &  =  & Q_{s,0}^2
   \frac{T_A(\bm{x}_\perp)}{T_{A,0}}
   \left(\frac{x_0}{x}\right)^\lambda
\end{eqnarray}
and similarly for a nucleus $B$.
Here, parameters $Q_{s,0}^2 = 2$ GeV$^2$, $T_{A,0}=1.53$ fm$^{-2}$,
$x_0=0.01$, and $\lambda=0.28$ are used.
Thickness function at each transverse coordinate
is obtained by counting the number of wounded nucleons $N$ within a tube
extending in the beam direction with radius
$r=\sqrt{\sigma_{NN}^{\text{in}}/\pi}$ from each grid point:
\begin{equation}
\label{eq:thickness}
T_A(\bm{x}_{\perp}) = \frac{N}{\sigma_{NN}^{\text{in}}}.
\end{equation}

For each generated configuration of nucleons
in colliding nuclei,
the $k_T$-factorization formula
is applied at each transverse coordinate
to obtain the distribution of produced gluons locally.
We apply the Kharzeev-Levin-Nardi (KLN) 
approach \cite{KLN01}
in the version previously employed in \cite{Hirano:2004rs}. 
In this approach, the distribution of gluons at
each transverse coordinate $\bm{x}_\perp$ produced with 
rapidity $y$ is given by the $k_T$-factorization formula \cite{GLR83}
\begin{eqnarray}
   \frac{dN_g}{d^2x_{\perp}dy}&=&
   \frac{2\pi^2}{C_F} \int\frac{d^2p_T}{p_T^2}
   \int^{p_T} \frac{d^2k_T}{4} \alpha_s(Q^2)   \nonumber\\
  &\times&      \phi_A(x_1,(\bm{p}_T{+}\bm{k}_T)^2/4;\bm{x}_\perp) \nonumber\\
  &\times&      \phi_B(x_2,(\bm{p}_T{-}\bm{k}_T)^2/4;\bm{x}_\perp), 
\label{eq:ktfac}
\end{eqnarray}
where $x_{1,2}{\eq}p_T\exp(\pm y)/\sqrt{s}$ and $p_T$ is the transverse 
momentum of the produced gluons. We choose an upper limit of 10\,GeV/$c$
for the $p_T$ integration. For the unintegrated gluon distribution 
function we use
\begin{equation}
\label{eq:uninteg}
  \phi_A(x,k^2_T;\bm{x}_\perp)=
\frac{\kappa C_F}{2\pi^3}
\frac{(1-x)^4}{\alpha_s(Q_s^2)}\frac{Q_s^2}{\max(Q_s^2,k^2_T)},
\end{equation}
where $C_F = (N_c^2-1)/(2N_c)$. The parameter $\kappa^2=1.75$
is chosen for the overall normalization of the gluon multiplicity
in order to fit the multiplicity data in Au+Au collisions 
at RHIC \cite{PHOBOS_Nch}.

As an initial condition for hydrodynamical calculations,
the initial entropy density in the transverse plane is obtained by
\begin{eqnarray}
s_0(\bm{x}_\perp) & = & 3.6 n_g \nonumber\\
& = & 3.6 \left. \frac{dN_g}{\tau_0 d^2 x_{\perp}d\eta_s}\right|_{y=\eta_s=0}.
\end{eqnarray}
Here we identify gluon's momentum rapidity $y$
with space-time rapidity $\eta_s$.

 \begin{figure*}[htb]
\includegraphics[width=3.4in]{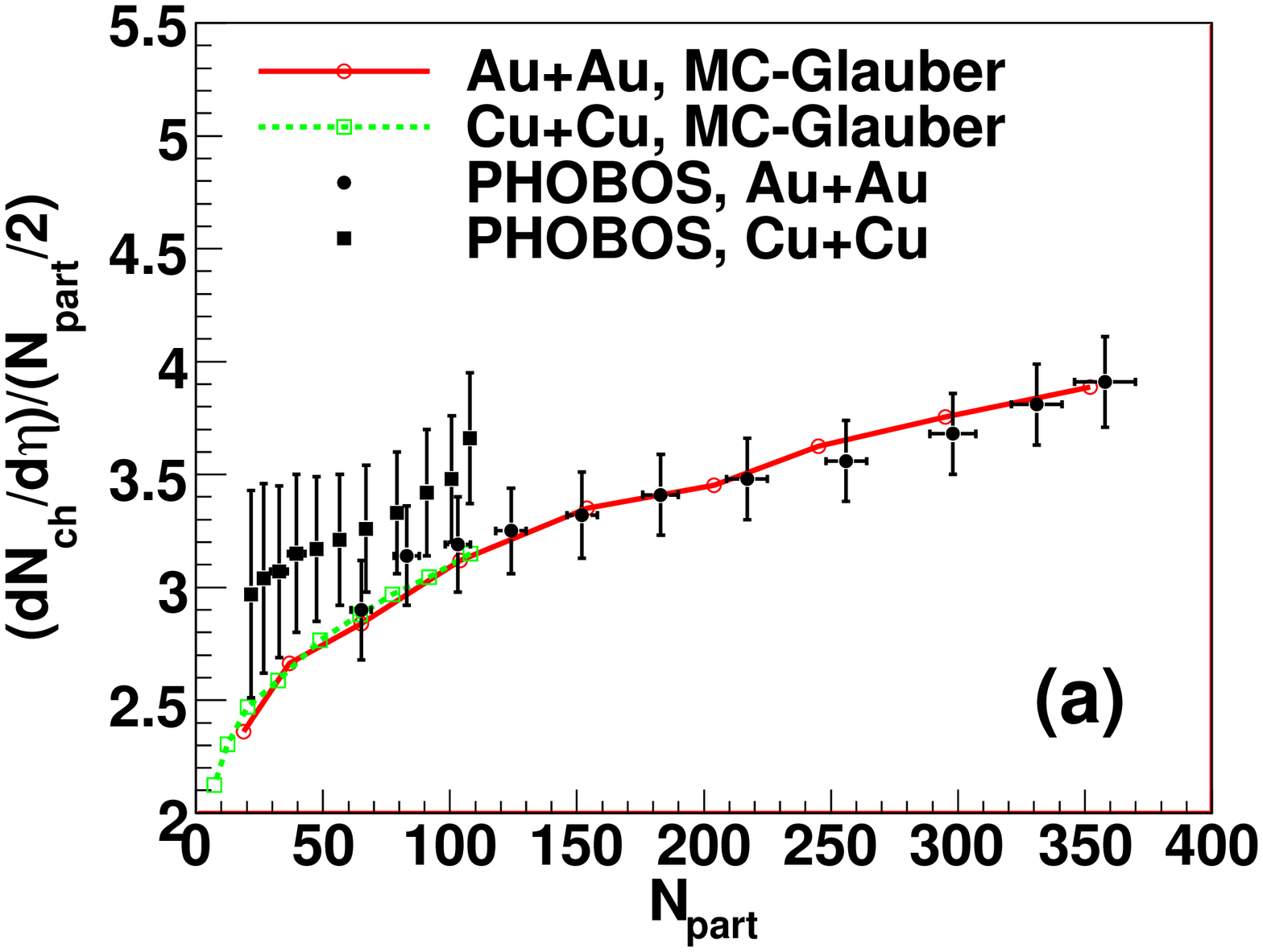}
\includegraphics[width=3.4in]{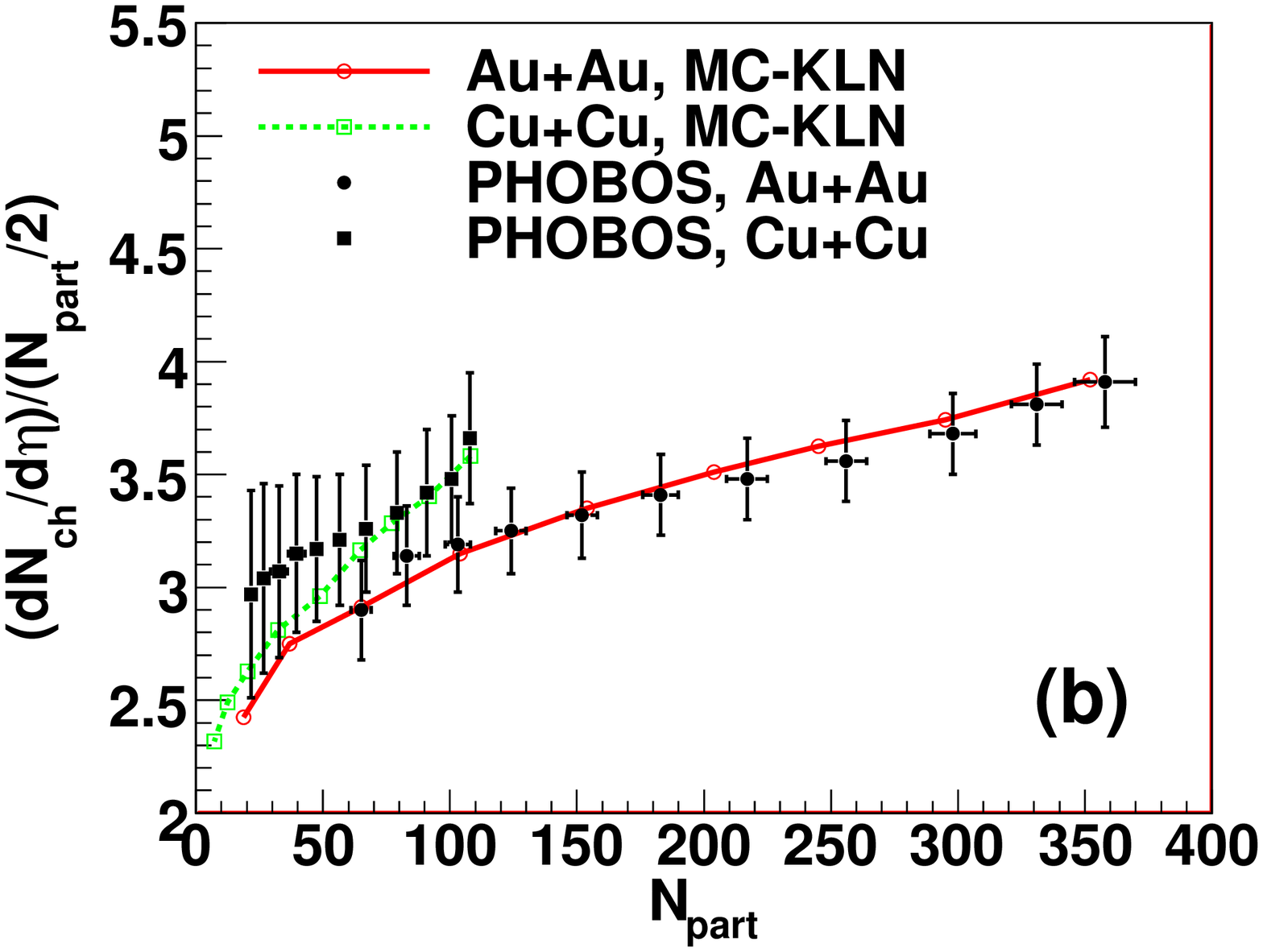}
\caption{(Color online)
Centrality dependences of charged particle multiplicity
in Au+Au (solid line) and Cu+Cu (dashed line) collisions
at $\sqrt{s_{NN}} = 200$ GeV.
(a) Initial conditions taken from the MC-Glauber model.
(b) Initial conditions taken from the MC-KLN model.
Experimental data are from Refs.~\cite{PHOBOS_Nch,Alver:2008ck}.
}
 \label{fig:dndeta}
 \end{figure*}
%

%
 \begin{figure*}[htb]
\includegraphics[width=3.4in]{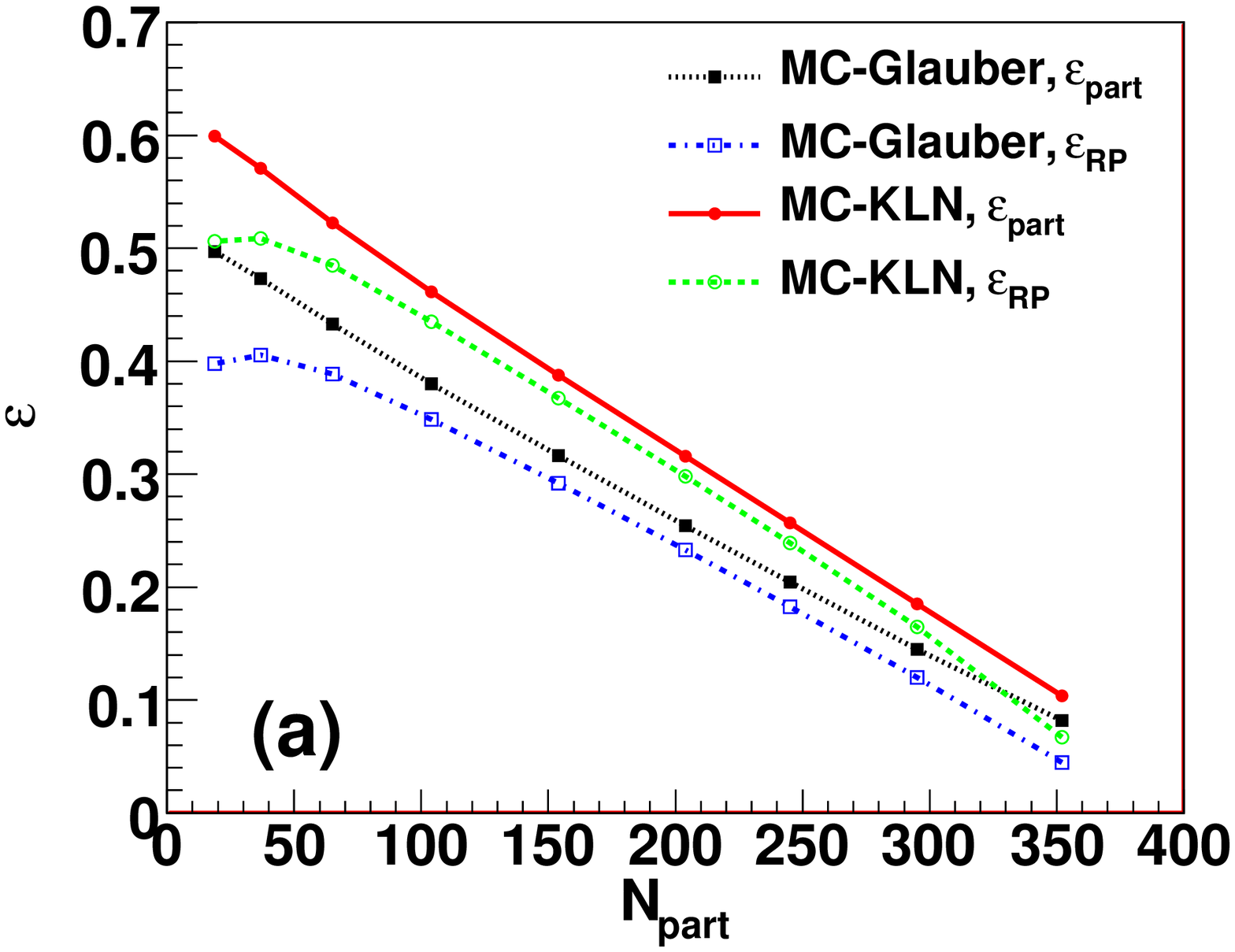}
\includegraphics[width=3.4in]{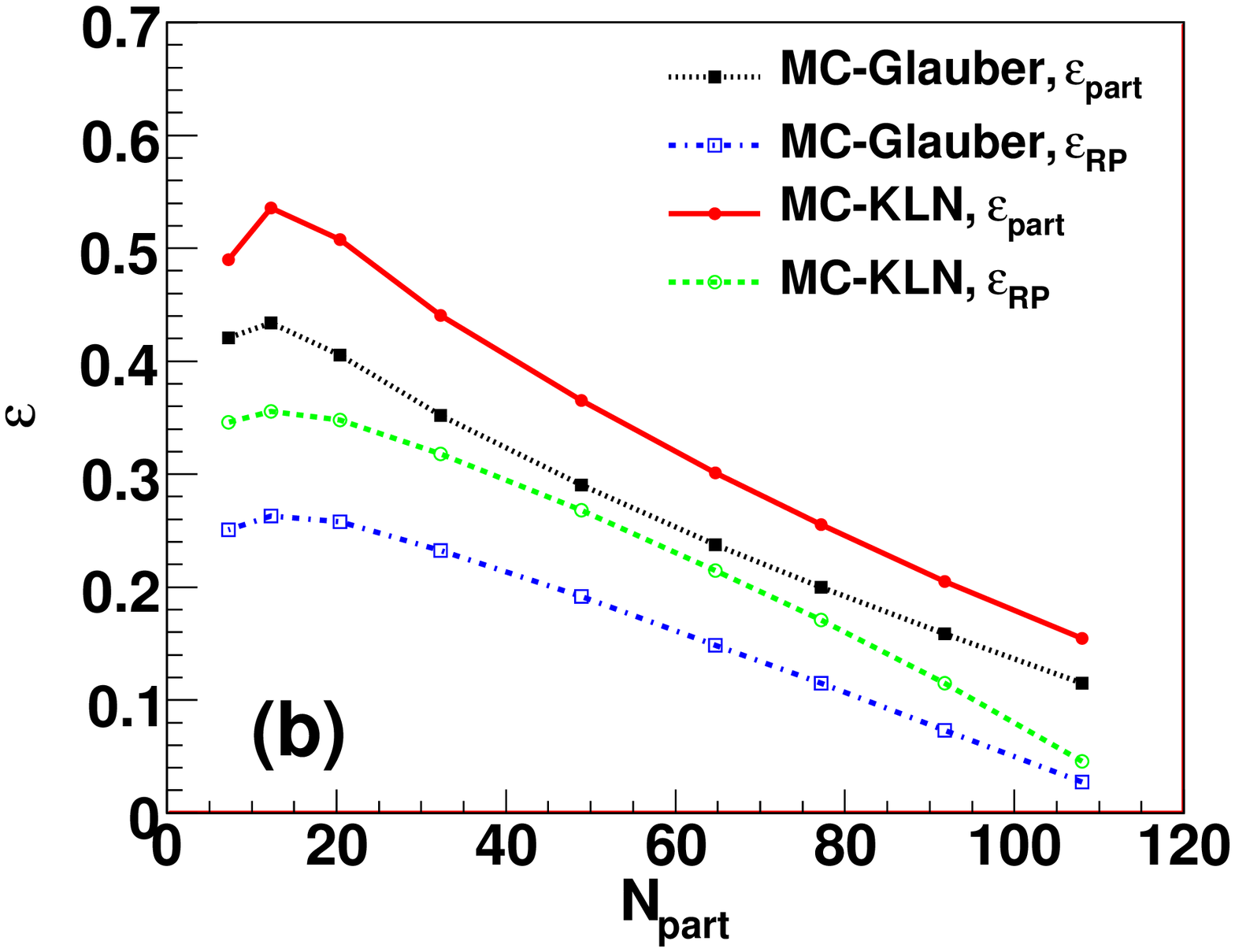}
\caption{(Color online)
Eccentricity as a function of $N_{\mathrm{part}}$
in (a) Au+Au and (b) Cu+Cu collisions.
Solid, dotted, dashed, and dash-dotted
lines correspond to
$\varepsilon_{\mathrm{part}}$ from the MC-KLN model,
$\varepsilon_{\mathrm{part}}$ from the MC-Glauber model,
$\varepsilon_{\mathrm{RP}}$ from the MC-KLN model,
and
$\varepsilon_{\mathrm{RP}}$ from the MC-Glauber model,
respectively.
}
 \label{fig:ecc}
 \end{figure*}
%

\section{Results}
\label{sec3}

\subsection{Centrality dependence of multiplicity}

Centrality dependences of multiplicity in the
MC-Glauber model
are compared with the PHOBOS data \cite{PHOBOS_Nch,Alver:2008ck}
in Fig.~\ref{fig:dndeta} (a).
Centrality dependence in Au+Au collisions is well reproduced
with a two component (soft + hard)
model with a small fraction of
hard component $\delta=0.14$.
The MC-KLN model with the setting mentioned
in the previous section also gives a reasonable agreement with
the data as shown in Fig.~\ref{fig:dndeta} (b).
We did not tune
initial parameters in Cu+Cu collisions.
The multiplicity in the Glauber model
depends only on $N_{\mathrm{part}}$
regardless of the collision system,
while a non-trivial $N_{\mathrm{part}}$ dependence
appears in the MC-KLN model:
At a fixed $N_{\mathrm{part}}$, 
$(dN_{\mathrm{ch}}/d\eta)/(N_{\mathrm{part}}/2)$
in Cu+Cu collisions is larger than the one
in Au+Au collisions.
All results are obtained by solving the hydrodynamic
equations below to $T_{\mathrm{dec}}=100$ MeV.
In the actual calculations of elliptic flow coefficients,
we replace the hadronic fluids
with the hadronic gases utilizing a hadronic cascade model.
Nevertheless, the centrality dependence of the multiplicity
is still within error bars \cite{HHKLN3}.

\subsection{Eccentricity}

The eccentricities as functions of $N_{\mathrm{part}}$
with or without eccentricity fluctuations
are compared in the MC-Glauber and MC-KLN models
in Au+Au (Fig.~\ref{fig:ecc} (a)) and Cu+Cu (Fig.~\ref{fig:ecc} (b))
collisions.
The impact parameter range,
the number of participants, the 
eccentricity, and the transverse area
are summarized 
in Appendix \ref{sec:appendix}
for each centrality from 0-5\% to 60-70\% 
in Au+Au (Table \ref{table:auau}) and Cu+Cu
(Table \ref{table:cucu}) collisions at $\sqrt{s_{NN}} = 200$ GeV.
In semicentral Au+Au collisions (10-50\% centrality),
the effect of initial fluctuations
enhances eccentricity parameter by 8-11\% (5-8\%)
in the MC-Glauber (MC-KLN) model.
The enhancement factor
$\varepsilon_{\mathrm{part}}/\varepsilon_{\mathrm{RP}}$
is the largest at the very central bin (0-5\% centrality):
$\varepsilon_{\mathrm{part}}/\varepsilon_{\mathrm{RP}} =$ 1.83
in the MC-Glauber model and 1.53 in the MC-KLN model.
A qualitatively similar behavior is observed
in Cu+Cu collisions as shown in Fig.~\ref{fig:ecc} (b).
However, it is quantitatively different: the enhancement factor 
is 1.51-1.74 (1.36-1.50) in the MC-Glauber
(MC-KLN) model in semicentral collisions (10-50\% centrality). 
In the very central events (0-5\% centrality),
the factor reaches 4.20 (3.36) in the MC-Glauber (MC-KLN) model.
Thus the resultant elliptic flow coefficient
is expected to be enhanced by almost the same amount of the
factor due to eccentricity fluctuations
especially in Cu+Cu collisions. We will demonstrate it
by using a dynamical model in the next subsection.

%
\begin{figure*}[htb]
\includegraphics[width=3.4in]{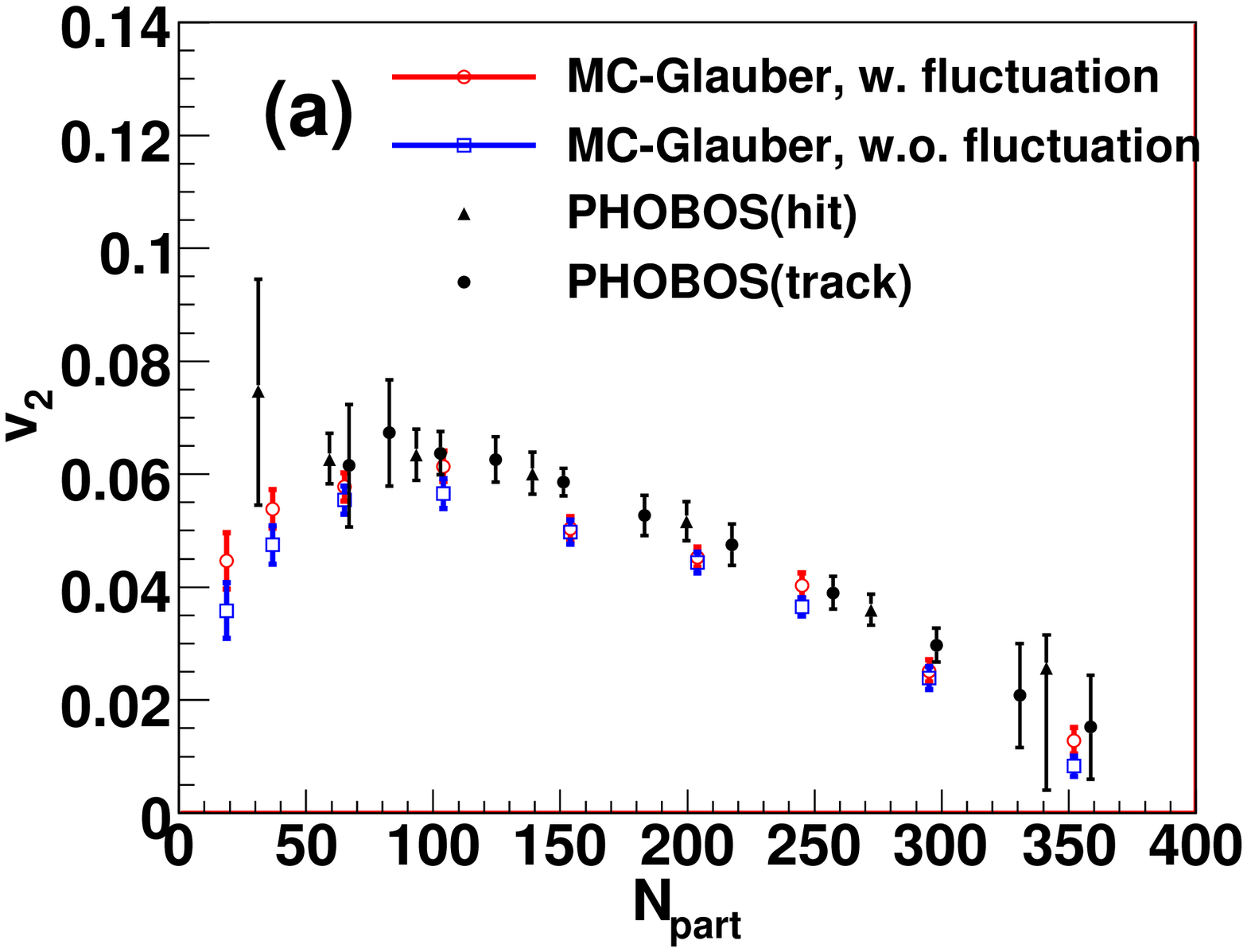}
\includegraphics[width=3.4in]{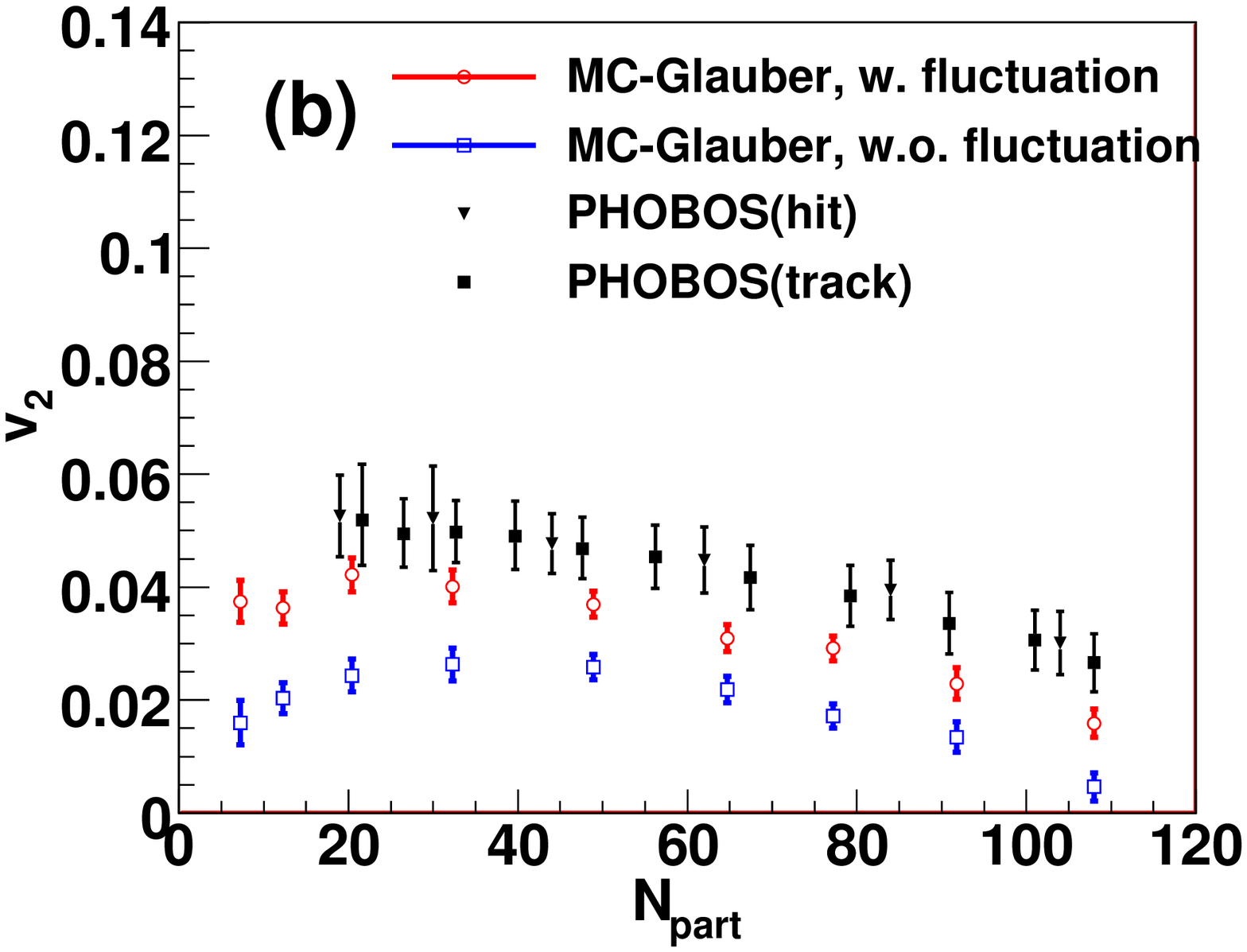}
\caption{(Color online)
Centrality dependences of $v_2$ for charged particles
at mid-rapidity in (a) Au+Au and (b) Cu+Cu collisions
at $\sqrt{s_{NN}} = 200$ GeV with the MC-Glauber model initial conditions
are compared with PHOBOS data (filled plots) \cite{Back:2004mh}.
Open circles (squares) are results with (without) eccentricity fluctuation.
}
 \label{fig:v2Glauber}
 \end{figure*}
%
%
\begin{figure*}[htb]
\includegraphics[width=3.4in]{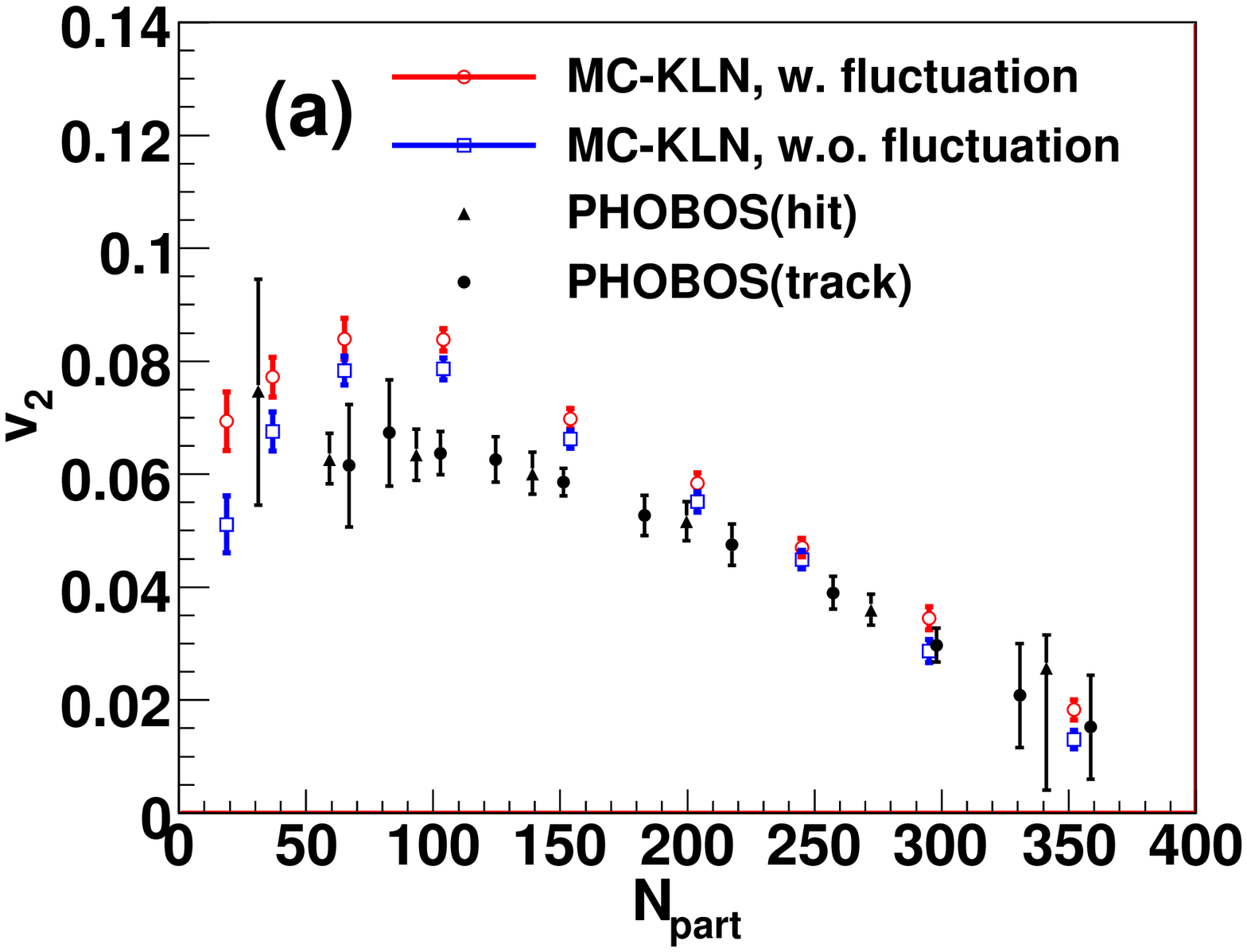}
\includegraphics[width=3.4in]{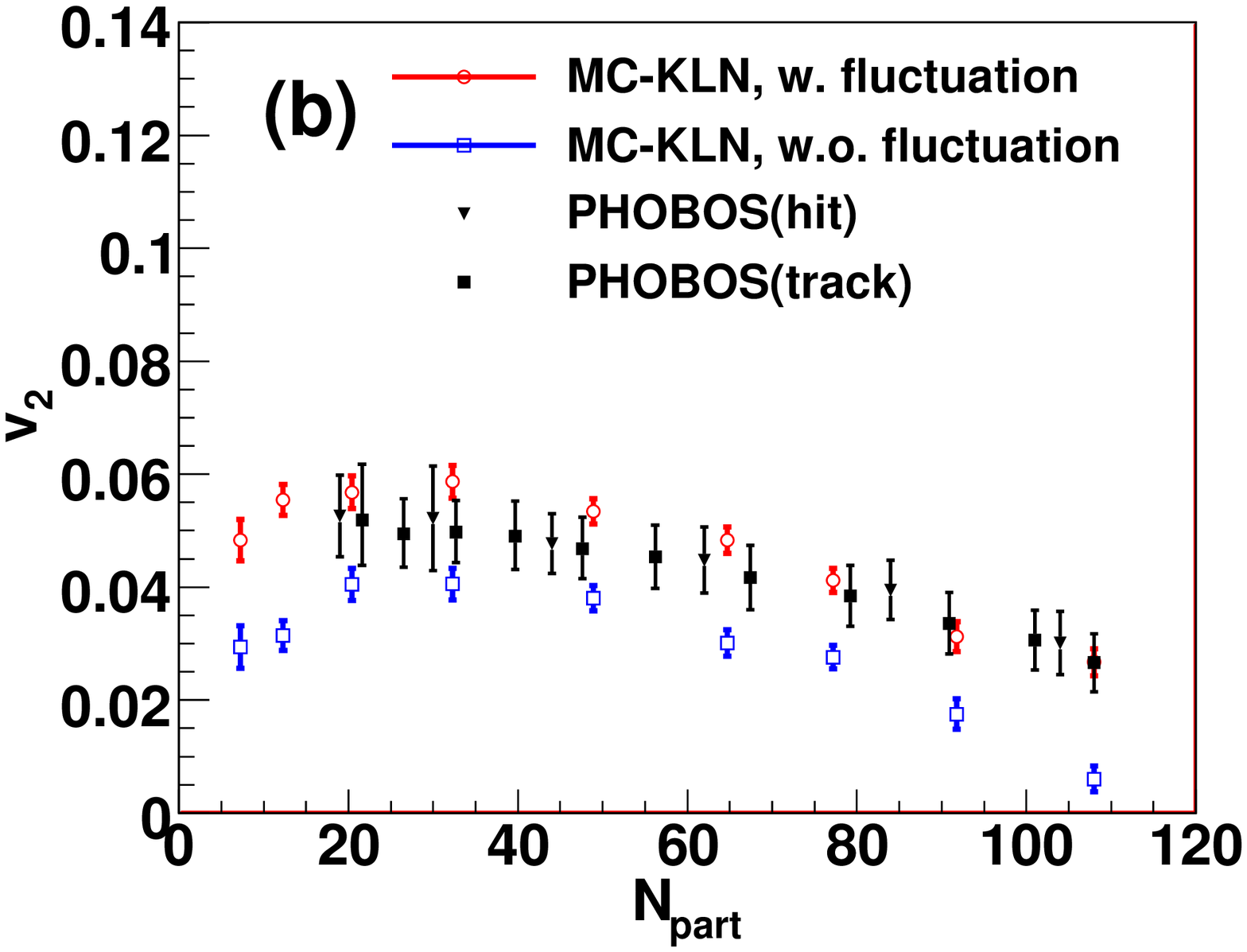}
\caption{(Color online)
Centrality dependences of $v_2$ for charged particles
at mid-rapidity in (a) Au+Au and (b) Cu+Cu collisions
at $\sqrt{s_{NN}} = 200$ GeV with the MC-KLN model initial conditions
are compared with PHOBOS data (filled plots) \cite{Back:2004mh}.
Open circles (squares) are results with (without) eccentricity fluctuation.
}
 \label{fig:v2kln}
 \end{figure*}
%

$\varepsilon_{\mathrm{RP}}$ in the MC-Glauber model
is consistent with our previous result
in which we used the standard eccentricity \cite{HHKLN}.
Notice that the consistency is
achieved only after retuning the Woods-Saxon parameters
due to finite nucleon profiles
discussed in the previous section.
$\varepsilon_{\mathrm{RP}}$ in the MC-KLN model
is slightly smaller \cite{Drescher:2006ca} than the result from
the naive KLN model \cite{HHKLN}.
In the MC-KLN model, there exists
the minimum saturation scale which is nothing but
the saturation scale of a nucleon.
On the other hand,
the saturation scale in the naive KLN model can be arbitrary small
below to $\Lambda_{\mathrm{QCD}}$,
which leads to
a sharper transverse profile and large eccentricity.
We found eccentricity 
is reduced by $\sim 15\%$
in the MC-KLN model compared to the KLN model, which would
affect the elliptic flow coefficients at the same amount.

\subsection{Elliptic flow coefficient}

Figure \ref{fig:v2Glauber}
shows the centrality dependence of
$v_2$ for charged hadrons
at mid-rapidity ($\mid \eta \mid < 1$)
in the Glauber model initialization
in (a) Au+Au and (b) Cu+Cu collisions at $\sqrt{s_{NN}} = 200$ GeV.
Experimental data \cite{Back:2004mh}
are reasonably reproduced
in Au+Au collisions. 
The effect of eccentricity fluctuations
is not significant in Au+Au collisions,
whereas
$v_2$ is largely enhanced 
in Cu+Cu collisions due to fluctuation effects.
These tendencies are also expected from the results of
initial eccentricity as shown in Fig.~\ref{fig:ecc}.
Interestingly elliptic flow coefficients 
$v_2$ in the Glauber model initial conditions still slightly undershoot
the experimental data, in particular, in Cu+Cu collisions
even with fluctuation effects.

Figure \ref{fig:v2kln} is the same as Fig.~\ref{fig:v2Glauber}
but the initial conditions are taken from the MC-KLN model.
Again, the effect of eccentricity fluctuations is small
in Au+Au collisions but is large in Cu+Cu collisions.
Due to larger initial eccentricity in the MC-KLN model
than the MC-Glauber model,
the results are somewhat larger than the experimental data
in peripheral Au+Au collisions.
Whereas, we reasonably reproduce
the $v_2$ data in Cu+Cu collisions.
It should be noted that 
the results in Au+Au collisions
lie systematically below the ones
in Fig.~2 of Ref.~\cite{HHKLN} which employed
the naive KLN model.
This is due to the reduction of eccentricity
by employing the fKLN model.

In both cases, the results using a more sophisticated
EOS such as the one from the 
numerical simulations of the lattice QCD \cite{Cheng:2007jq}
would be demanded.
The EOS from the lattice simulations
has a cross-over behavior rather than the first order
phase transition.
Therefore it is  harder in the vicinity of pseudo-critical temperature
$T_{c} \sim 190$ MeV
and softer in the high temperature region up to $T \sim 2T_{c}$
than the one employed in the present study.
Because of the above reasons, whether final $v_2$ is enhanced or reduced
in comparison with the current results
would be non-trivial in the case of the EOS from the lattice QCD.
Detailed studies
on the dependences of $v_2$
on the EOS
will be reported elsewhere.

\section{Conclusions}
\label{sec4}

We calculated the elliptic flow
parameter as a function of the number of participants
in the QGP hydro plus the hadronic cascade model
and found that the effect of eccentricity fluctuations is 
visible in very central and peripheral Au+Au collisions and 
is quite large in Cu+Cu collisions.
This strongly suggests that the effect of eccentricity fluctuations
is an important factor which has to be included in the dynamical model
for understanding of the elliptic flow data
and for precise extraction
of transport properties from the data.

We also found that a finite
nucleon size assumed in the conventional Monte Carlo
approaches reduces initial eccentricity
by $\sim$10\% with a default Woods-Saxon
parameter set.
This requires reparametrization of
the nuclear radius and the diffuseness
parameter to obtain
the actual nuclear distribution
in the case of the finite profile of nucleons
as implemented in the Monte Carlo Glauber model.

In the case of the Glauber type initialization,
the results still undershot the experimental data a little
in both Au+Au and Cu+Cu collisions
even after inclusion of eccentricity fluctuation effects.
If one implemented the viscosity in the QGP phase,
the results of $v_2$ would get reduced.
So there is almost no room for the viscosity 
in the QGP stage to play a role
in the Glauber initial conditions within
the hybrid approach with the ideal gas EOS in the QGP phase.

On the other hand, 
we overpredicted $v_2$ in peripheral ($N_{\mathrm{part}}<150$)
Au+Au collisions in the CGC model.
Viscous effects in the QGP phase
could reduce the $v_2$ and
enable us to reproduce
the data in Au+Au collisions
in this case.
However, the results
are already comparable with
the data in Cu+Cu collisions
even though the number of participants is almost the same
as that in peripheral Au+Au collisions.
So it would be non-trivial whether
the same viscous effects 
also give the right amount of $v_2$
in Cu+Cu collisions.

So far, one has been focusing on comparison of hydrodynamic results
with $v_2$ data only in Au+Au collisions.
The experimental data in Cu+Cu collisions
also have a strong power to constrain the 
dynamical models. Therefore,
simultaneous analysis of 
$v_2$ data in both Au+Au and Cu+Cu collisions
will be called for in future
hydrodynamic studies.

\acknowledgments
We would like to thank Adrian Dumitru and Akira Ohnishi 
for fruitful discussions.
This work was initiated at
the workshop
``New Frontiers in QCD 2008"
under the project 
``Yukawa International Program for
Quark-Hadron Science" 
at Yukawa Institute for Theoretical Physics (YITP),
Kyoto University.
A part of work
was also done during the workshop
on ``Entropy production before QGP" at YITP.
The authors thank the organizers
of these workshops and members of YITP
for invitation and kind hospitality.
The work of T.H. was partly supported by
Grant-in-Aid for Scientific Research
No.~19740130 and by Sumitomo Foundation
No.~080734.  The work of Y.N. was supported by Grand-in-Aid for Scientific
Research No.~20540276.

\appendix

\section{Centrality dependence of eccentricity and transverse area}
\label{sec:appendix}

We obtain centrality in our model as follows.
Using the Glauber model, one calculates centrality, namely
the fraction of inelastic cross section as a function of impact parameter,
\begin{eqnarray}
\label{eq:cent}
c(b) & = & \frac{2 \pi \int_0^{b} b'db'\left(1-e^{-\sigma_{NN}^{\mathrm{in}}T_{AA}(b')}\right)}{2 \pi \int_0^{\infty} b'db'\left(1-e^{-\sigma_{NN}^{\mathrm{in}}T_{AA}(b')}\right)},\\
T_{AA}(b) & = &\int d^2x_\perp T_{A}\left(x-\frac{b}{2}, y \right)T_{A}\left(x+\frac{b}{2}, y \right).\nonumber \\
\end{eqnarray}
For a given range of centrality $c_{\mathrm{min}}<c<c_{\mathrm{max}}$,
the maximum and 
minimum impact parameters can be defined from Eq.~(\ref{eq:cent})
as $c(b_{\mathrm{max}}) = c_{\mathrm{max}}$ and
$c(b_{\mathrm{min}}) = c_{\mathrm{min}}$, respectively.
In generating initial profiles,
an impact parameter is randomly
chosen 
in $b_{\mathrm{min}}<b<b_{\mathrm{max}}$
for each centrality bin
with a probability 
\begin{equation}
P(b)db = \frac{2b}{b_{\mathrm{max}}^2-b_{\mathrm{min}}^2}db.
\end{equation}

Impact parameter ranges, the resultant number of participants,
eccentricity with respect to the reaction plane, participant eccentricity,
and transverse area are summarized in Table \ref{table:auau} 
for Au+Au collisions and Table \ref{table:cucu} for Cu+Cu collisions.
These parameters would be very useful
to see whether 
$v_2$ data reach the so-called hydrodynamic limit
or not \cite{Drescher:2007cd}.
It might be possible to divide all events
into centralities according to
the Monte Carlo results of, \textit{e.g.},
the $N_{\mathrm{part}}$ distribution.

\begin{table*}[h]
\caption{Centrality dependence of eccentricity
and transverse area in Au+Au collisions.}
\label{table:auau}
\begin{ruledtabular}
\begin{tabular}{cccccccccc}
Centrality(\%) & 0-5 & 5-10 & 10-15 & 15-20 & 20-30 & 30-40 & 40-50 & 50-60 & 60-70 \\
\hline
$b_{\mathrm{min}}$ (fm) & 0.0 & 3.3 & 4.7 & 5.8 & 6.7 & 8.2 & 9.4 & 10.6 & 11.6 \\
$b_{\mathrm{max}}$ (fm) & 3.3 & 4.7 & 5.8 & 6.7 & 8.2 & 9.4 & 10.6 & 11.6 & 12.5\\
$ N_{\mathrm{part}}$  & 352 & 295 & 245 & 204 & 154 & 104 & 65.1 & 36.8 & 18.8 \\
$\epsilon_{\mathrm{RP}}^{\mathrm{MC-Glauber}}$ & 0.0446 & 0.120 & 0.183 & 0.233 & 0.292 & 0.348 & 0.389& 0.405& 0.398\\
$\epsilon_{\mathrm{part}}^{\mathrm{MC-Glauber}}$ & 0.0818  & 0.145 & 0.204 & 0.254 & 0.316& 0.380 & 0.433& 0.473& 0.497 \\
$ S_{\mathrm{RP}}^{\mathrm{MC-Glauber}}$
(fm$^2$) & 23.4 & 20.5 & 18.0 & 16.0 & 13.5 & 10.9 & 8.69 & 6.78 & 5.07\\
$ S_{\mathrm{part}}^{\mathrm{MC-Glauber}}$ 
(fm$^2$) & 23.4  & 20.5 & 18.0 & 16.0 & 13.5 & 10.9 & 8.65 & 6.73 & 5.05 \\
$ \epsilon_{\mathrm{RP}}^{\mathrm{MC-KLN}}$ & 0.0671& 0.165 & 0.239& 0.298& 0.367& 0.435 & 0.485& 0.509 & 0.506 \\
$ \epsilon_{\mathrm{part}}^{\mathrm{MC-KLN}}$ & 0.103& 0.185 & 0.257 & 0.316 & 0.387& 0.461& 0.522& 0.571 & 0.599\\
$ S_{\mathrm{RP}}^{\mathrm{MC-KLN}}$ 
(fm$^2$) & 23.8 & 20.1 & 17.3 & 15.1 &  12.4 & 9.67 & 7.42& 5.55 &3.96\\
$ S_{\mathrm{part}}^{\mathrm{MC-KLN}}$
(fm$^2$) & 23.7  & 20.1 & 17.2 & 15.0 & 12.3& 9.55 & 7.26 & 5.33 & 3.67
\end{tabular}
\end{ruledtabular}
\end{table*}

\begin{table*}[h]
\caption{Centrality dependence of eccentricity
and transverse area in Cu+Cu collisions.}
\label{table:cucu}
\begin{ruledtabular}
\begin{tabular}{cccccccccc}
Centrality(\%) & 0-5 & 5-10 & 10-15 & 15-20 & 20-30 & 30-40 & 40-50 & 50-60 & 60-70 \\
\hline
$b_{\mathrm{min}}$ (fm) & 0.0 & 2.4 & 3.3 & 4.1 & 4.7 & 5.8 & 6.7 & 7.5 & 8.2 \\
$b_{\mathrm{max}}$ (fm) & 2.4 & 3.3 & 4.1 & 4.7 & 5.8 & 6.7 & 7.5 & 8.2 & 8.9  \\
$ N_{\mathrm{part}}$  & 108 & 91.8 & 77.2 & 64.7 & 48.9 & 32.3 & 20.4 & 12.3 & 7.27 \\
$ \epsilon_{\mathrm{RP}}^{\mathrm{MC-Glauber}}$ & 0.0274 & 0.0733 & 0.115  & 0.148 & 0.192 & 0.232 & 0.257 & 0.263 & 0.251\\
$ \epsilon_{\mathrm{part}}^{\mathrm{MC-Glauber}}$ & 0.115  & 0.159 & 0.200 & 0.237 & 0.290 & 0.352 & 0.406 & 0.434 & 0.421 \\
$ S_{\mathrm{RP}}^{\mathrm{MC-Glauber}}$ (fm$^2$)   & 12.1  & 11.0 & 10.1  & 9.15  & 7.98  & 6.57  & 5.32  & 4.12 & 3.00 \\
$ S_{\mathrm{part}}^{\mathrm{MC-Glauber}}$ (fm$^2$) & 12.1  & 11.0 & 10.0  & 9.12  & 7.94  & 6.54  & 5.30  & 4.14 & 3.11 \\
$ \epsilon_{\mathrm{RP}}^{\mathrm{MC-KLN}}$ & 0.0457& 0.115 & 0.171& 0.214 & 0.268 & 0.318 & 0.348& 0.356 & 0.346 \\
$ \epsilon_{\mathrm{part}}^{\mathrm{MC-KLN}}$ & 0.154& 0.205 & 0.256 & 0.301 & 0.365& 0.441& 0.508 & 0.536 & 0.490\\
$ S_{\mathrm{RP}}^{\mathrm{MC-KLN}}$ (fm$^2$) & 11.7& 10.3  & 9.18 & 8.18  & 6.96   & 5.54& 4.33 & 3.20 & 2.16\\
$ S_{\mathrm{part}}^{\mathrm{MC-KLN}}$ (fm$^2$) & 11.6& 10.2& 9.06 & 8.04  & 6.78    & 5.31& 4.04& 2.88 & 1.89
\end{tabular}
\end{ruledtabular}
\end{table*}



\begin{thebibliography}{99}

\bibitem{Ollitrault}
  J.~Y.~Ollitrault,
  Phys.\ Rev.\ D {\bf 46}, 229 (1992).

\bibitem{STARv2}
C.~Adler {\it et al.} [STAR Collaboration], Phys. Rev. Lett. {\bf 87}, 
182301 (2001);
J.~Adams {\it et al.} [STAR Collaboration], Phys. Rev. Lett. {\bf 92}, 
052302 (2001).

\bibitem{PHENIXv2}
K. Adcox {\it et al.} [PHENIX Collaboration], 
Phys. Rev. Lett. {\bf 89}, 212301 (2002);  
S.~S.~Adler {\it et al.} [PHENIX Collaboration],
Phys. Rev. Lett. {\bf 91}, 182301 (2003).


\bibitem{reviews}
See also recent reviews
from theoretical and experimental points of view:
  P.~Huovinen and P.~V.~Ruuskanen,
  Ann.\ Rev.\ Nucl.\ Part.\ Sci.\  {\bf 56}, 163 (2006);
  J.~Y.~Ollitrault,
  Eur.\ J.\ Phys.\  {\bf 29}, 275 (2008);
  T.~Hirano, N.~van der Kolk, and A.~Bilandzic,
  arXiv:0808.2684 [nucl-th];
  S.~A.~Voloshin, A.~M.~Poskanzer and R.~Snellings,
  arXiv:0809.2949 [nucl-ex];
  U.~W.~Heinz,
  arXiv:0901.4355 [nucl-th];
  P.~Romatschke,
  arXiv:0902.3663 [hep-ph].


\bibitem{Song}
  H.~Song and U.~W.~Heinz,
  Phys.\ Lett.\  B {\bf 658}, 279 (2008);
  Phys.\ Rev.\  C {\bf 77}, 064901 (2008);
  Phys.\ Rev.\ C {\bf 78}, 024902 (2008).


\bibitem{Chaudhuri}
A.~K.~Chaudhuri,
  arXiv:0704.0134 [nucl-th];
  arXiv:0708.1252 [nucl-th];
  arXiv:0801.3180 [nucl-th];
  Phys.\ Lett.\  B {\bf 672}, 126 (2009).

\bibitem{Baier}
  P.~Romatschke and U.~Romatschke,
  Phys.\ Rev.\ Lett.\  {\bf 99}, 172301 (2007);
  M.~Luzum and P.~Romatschke,
  Phys.\ Rev.\  C {\bf 78}, 034915 (2008);
  arXiv:0901.4588 [nucl-th].
  

\bibitem{Dusling}
  K.~Dusling and D.~Teaney,
  Phys.\ Rev.\  C {\bf 77}, 034905 (2008).

\bibitem{Denicol}
G.~Denicol, talk at Quark Matter 2009, 
Knoxville, USA, March 30-April 4, 2009.  
  
\bibitem{Teaney}
  D.~Teaney,
  Phys.\ Rev.\  C {\bf 68}, 034913 (2003).

\bibitem{Monnai}
  A.~Monnai and T.~Hirano,
  arXiv:0903.4436 [nucl-th].


\bibitem{HG05}
  T.~Hirano and M.~Gyulassy,
  Nucl.\ Phys.\ A {\bf 769}, 71 (2006).

\bibitem{HHKLN}
  T.~Hirano, U.~Heinz, D.~Kharzeev, R.~Lacey, and Y.~Nara,
  Phys.\ Lett.\ B {\bf 636}, 299 (2006).

\bibitem{Kuhlman:2006qp} 
  A.~Kuhlman, U.~Heinz, and Y.~V.~Kovchegov, 
  Phys.\ Lett.\ B {\bf 638}, 171 (2006);
  A.~Adil, H.~J.~Drescher, A.~Dumitru, A.~Hayashigaki, and Y.~Nara, 
  Phys.\ Rev.\  C {\bf 74}, 044905 (2006);
  T.~Lappi and R.~Venugopalan, 
  Phys.\ Rev.\  C {\bf 74}, 054905 (2006).



\bibitem{MS03}
  M.~Miller and R.~Snellings,
  nucl-ex/0312008.

\bibitem{Zhu:2005qa}
  X.~l.~Zhu, M.~Bleicher and H.~Stoecker, 
  Phys.\ Rev.\ C {\bf 72}, 064911 (2005).

\bibitem{Bhalerao:2006tp}
  R.~S.~Bhalerao and J.~Y.~Ollitrault,
  Phys.\ Lett.\  B {\bf 641}, 260 (2006).


\bibitem{phobosfluc}
  B.~Alver {\it et al.} [PHOBOS Collaboration], 
  nucl-ex/0702036; 
  Phys.\ Rev.\  C {\bf 77}, 014906 (2008).

\bibitem{Andrade:2006yh}
  R.~Andrade, F.~Grassi, Y.~Hama, T.~Kodama, and O.~Socolowski Jr., 
  Phys.\ Rev.\ Lett.\  {\bf 97}, 202302 (2006).

\bibitem{Drescher:2006ca}
  H.~J.~Drescher and Y.~Nara,
  Phys.\ Rev.\  C {\bf 75}, 034905 (2007);
  Phys.\ Rev.\  C {\bf 76}, 041903 (2007).

\bibitem{Broniowski:2007ft}
  W.~Broniowski, P.~Bozek, and M.~Rybczynski,
  Phys.\ Rev.\  C {\bf 76}, 054905 (2007).

\bibitem{Broniowski:2007nz}
  W.~Broniowski, M.~Rybczynski, and P.~Bozek,
  Comput.\ Phys.\ Commun.\  {\bf 180}, 69 (2009).

\bibitem{Voloshin:2007pc}
  S.~A.~Voloshin, A.~M.~Poskanzer, A.~Tang, and G.~Wang,
  Phys.\ Lett.\  B {\bf 659}, 537 (2008).


\bibitem{HHKLN2}
T.~Hirano, U.~Heinz, D.~Kharzeev, R.~Lacey, and Y.~Nara,
  Phys.\ Rev.\  C {\bf 77}, 044909 (2008).

\bibitem{BassDumitru}
  A.~Dumitru, S.~A.~Bass, M.~Bleicher, H.~St\"ocker, and W.~Greiner,
  Phys.\ Lett.\ B {\bf 460}, 411 (1999);
  S.~A.~Bass, A.~Dumitru, M.~Bleicher, L.~Bravina, E.~Zabrodin, 
  H.~St\"ocker, and W.~Greiner,
  Phys.\ Rev.\ C {\bf 60}, 021902 (1999);
  S.~A.~Bass and A.~Dumitru,
  Phys.\ Rev.\ C {\bf 61}, 064909 (2000).

\bibitem{TLS01}
  D.~Teaney, J.~Lauret, and E.V.~Shuryak,
  Phys.\ Rev.\ Lett.\  {\bf 86}, 4783 (2001);
  nucl-th/0110037.

\bibitem{NonakaBass}
  C.~Nonaka and S.~A.~Bass,
  Nucl.\ Phys.\  A {\bf 774}, 873 (2006);
  Phys.\ Rev.\  C {\bf 75}, 014902 (2007).

\bibitem{Petersen:2008dd}
  H.~Petersen, J.~Steinheimer, G.~Burau, M.~Bleicher, and H.~Stocker,
  Phys.\ Rev.\  C {\bf 78}, 044901 (2008);
  H.~Petersen and M.~Bleicher,
  arXiv:0901.3821 [nucl-th].

\bibitem{Werner}
K.~Werner, T.~Hirano, Iu Karpenko, T.~Pierog, S.~Porteboeuf,
M.~Bleicher, and S.~Haussler, J.~Phys.~G: Nucl.~Part.~Phys. (in press).

\bibitem{CF}
  F.~Cooper and G.~Frye,
  Phys.\ Rev.\ D {\bf 10}, 186 (1974).

\bibitem{BRAHMS:stopping}
I.~G.~Bearden \textit{et al.}, [BRAHMS Collaboration],
Phys. Rev. Lett. \textbf{93}, 102301 (2004).

\bibitem{Hiranov2eta}
  T.~Hirano,
  Phys.\ Rev.\ C {\bf 65}, 011901 (2002).

\bibitem{HT02}
  T.~Hirano and K.~Tsuda,
  Phys.\ Rev.\ C {\bf 66}, 054905 (2002).

\bibitem{Bebie}
  H.~Bebie, P.~Gerber, J.~L.~Goity, and H.~Leutwyler,
  Nucl.\ Phys.\  B {\bf 378}, 95 (1992).

\bibitem{spherio}
  N.~Arbex, F.~Grassi, Y.~Hama, and O.~Socolowski Jr.,
  Phys.~Rev.~C {\bf 64}, 064906 (2001);
  W.~L.~Qian, R.~Andrade, F.~Grassi, O.~Socolowski Jr., T.~Kodama, and Y.~Hama,
  Int.\ J.\ Mod.\ Phys.\  E {\bf 16}, 1877 (2007).

\bibitem{Teaney:2002aj}
  D.~Teaney,
  nucl-th/0204023.

\bibitem{Kolb:2002ve}
  P.~F.~Kolb and R.~Rapp,
  Phys.\ Rev.\  C {\bf 67}, 044903 (2003).

\bibitem{Huovinen:2007xh}
  P.~Huovinen,
  arXiv:0710.4379 [nucl-th].

\bibitem{jam} 
  Y.~Nara, N.~Otuka, A.~Ohnishi, K.~Niita, and S.~Chiba,
  Phys. Rev. C {\bf 61}, 024901 (2000).


\bibitem{Andrade:2008xh}
  R.~P.~G.~Andrade, F.~Grassi, Y.~Hama, T.~Kodama, and W.~L.~Qian,
  Phys.\ Rev.\ Lett.\  {\bf 101}, 112301 (2008).

\bibitem{Miller:2007ri}
  M.~L.~Miller, K.~Reygers, S.~J.~Sanders, and P.~Steinberg,
  Ann.\ Rev.\ Nucl.\ Part.\ Sci.\  {\bf 57}, 205 (2007).




\bibitem{KLN01}
  D.~Kharzeev and M.~Nardi, Phys.\ Lett.\ B {\bf 507}, 121 (2001);
  D.~Kharzeev and E.~Levin, Phys.\ Lett.\ B {\bf 523}, 79 (2001);
  D.~Kharzeev, E.~Levin, and M.~Nardi, Phys.\ Rev.\ C {\bf 71}, 054903 (2005);
  D.~Kharzeev, E.~Levin, and M.~Nardi, Nucl.\ Phys.\ A {\bf 730}, 448 (2004).

\bibitem{Hirano:2004rs} 
  T.~Hirano and Y.~Nara, 
  Nucl.\ Phys.\ A {\bf 743}, 305 (2004).

\bibitem{De Jager:1987qc}
  H.~De Vries, C.~W.~De Jager, and C.~De Vries,
  Atom.\ Data Nucl.\ Data Tabl.\  {\bf 36}, 495 (1987).

\bibitem{PHOBOS_Nch}
  B.~B.~Back {\it et al.} [PHOBOS Collaboration], 
  Phys. Rev. C {\bf 65}, 061901 (2002).



\bibitem{GLR83}
  L.~V.~Gribov, E.~M.~Levin, and M.~G.~Ryskin,
  Phys.\ Rept.\  {\bf 100}, 1 (1983).




\bibitem{Alver:2008ck}
  B.~Alver {\it et al.}  [PHOBOS Collaboration],
  arXiv:0808.1895 [nucl-ex].

\bibitem{HHKLN3}
T.~Hirano, U.~Heinz, D.~Kharzeev, R.~Lacey, and Y.~Nara,
  J.\ Phys.\ G: Nucl.~Part.~Phys.~{\bf 34}, S879 (2007).



\bibitem{Back:2004mh}
  B.~B.~Back {\it et al.}  [PHOBOS Collaboration],
  Phys.\ Rev.\  C {\bf 72}, 051901 (2005);
  B.~Alver {\it et al.}  [PHOBOS Collaboration],
  Phys.\ Rev.\ Lett.\  {\bf 98}, 242302 (2007).

\bibitem{Cheng:2007jq}
  M.~Cheng {\it et al.},
  Phys.\ Rev.\  D {\bf 77}, 014511 (2008);
  A.~Bazavov {\it et al.},
  arXiv:0903.4379 [hep-lat].

\bibitem{Drescher:2007cd}
 H.~J.~Drescher, A.~Dumitru, C.~Gombeaud, and J.~Y.~Ollitrault,
 Phys.\ Rev.\  C {\bf 76}, 024905 (2007).

  
\end{thebibliography}

\end{document}